\documentstyle[12pt]{article}

\advance \textheight by \topskip
\begin{document}

\begin{center} {\large\bf 
Gamma-Ray Bursts: Super-Explosions in the Universe and Related High-Energy
Phenomena}\\
\bigskip
	
{\bf K.A. Postnov$^{1,2}$}\\

{\it $^1$ Faculty of Physics, Moscow State University, 119899 Moscow, Russia\\

$^2$ Sternberg Astronomical Institute,
119899 Moscow, Russia}

\end {center}
\bigskip

{\it Abstract}. The recent  progress in studies of gamma-ray bursts, their
afterglows, and host galaxies is discussed. The emphasis is given to
high-energy phenomena associated with gamma-ray burst explosions:
high-energy cosmic rays, neutrinos, gravitational waves. We also show how
the relativistic fireball model for GRBs can be used to constrain modern
theories of large and infinite extra-dimensions. In particular, in the frame
of 5D gravity with the Standard Model localized on 3D brane (Dvali et al.
2000), the very existence of relativistic fireballs of $\sim 10^{53}$ ergs
puts the lower bound on the quantum gravity scale $\sim 0.1$ eV.

\section{Introduction and short history}

A large progress in cosmic gamma-ray bursts (GRB) have been made in 
recent years, after their precise localizations on the sky have become
possible with observations of their soft afterglows.
There is a number of 
modern reviews devoted to the problem of GRB in general
(e.g., Piran 1999,2001, Postnov 1999, Meszaros and Rees 2001), and we 
here we will not discuss the GRB problem in full detail.
Instead, we will try to present information most relevant to the
topics of this School.

We start with very brief history of GRB studies, which clearly can be
subdivided into three periods. The first period started after serendipitous
discovery of cosmic GRB by american military {\it Vela} satellites in the
end of 60s and ended in 1992 with the launch of Compton Gamma-Ray
Observatory. This was a period of data accumulation, with the major
contribution made by {\it Konus} experiment (Mazets et al. 1981).
As there are no means to measure a distance to a poorly localized 
GRB, it was completely unclear at that time if GRB were 
local, galactic or remote extragalactic events. 

The BATSE era continued until the beginning of 1997, when the first 
X-ray afterglow from GRB970228 was detected by 
the Dutch-Italian Beppo-SAX satellite (Costa et al. 1997).
Before Beppo-SAX, 
poor localization (of order degrees) of GRB positions on the sky has prevented 
their full astronomical investigation from being made, and most 
valuable BATSE contribution has been the accumulation of a large homogeneous
collection of GRBs (more than three thousand), which allowed  thorough
statistical studies of these objects (Paciesas et al. 1999).
These studies (especially, log N-log S counts) provided an indirect
evidence that we are dealing with extragalactic events, located 
at gigaparsec distance scale, so their energetics must be unusually
high, of order of $10^{52}$ ergs on average.     
These clues have been confirmed by the identification of the 
location of more than a dozen GRB afterglows inside high-redshift galaxies,
which firmly established the cosmological nature of most cosmic GRBs
\footnote{Short (< 2 s) GRBs are still unidentified in other wavelengths and
might represent a separate phenomenon, however their 
spectral characteristics are very similar to those of long cosmological GRBs
(Frederiks et al. 2001)}. 

At present, 17 reliable redshift determinations of GRBs or their
hosts are known (see Bloom et al. 2001),
more or less uniformly distributed between $z=0.43$ and $z=4.5$
\footnote{We will not consider still controversial case of GRB980425
possibly associated with supernova explosion SN1998bw in a nearby galaxy.}

\section{Relativistic fireball model}

Of more than 200 GRB models of the mid-80s, the most viable one proved 
to be the relativistic
fireball model, which seems to be confirmed by the bulk of GRB studies in a wide
range of wavelengths from radio to gamma-rays (see Piran 1999, 2001 for a
comprehensive review) (for an alternative explanation 
see for example Ruffini et al. 2001). 
A huge energy ($\Delta E\sim 10^{51}-10^{53}$ ergs)
in gamma-rays ($E_\gamma\sim 100$ keV -- 10 MeV), released in a short
observed duration of GRBs (typically, $\Delta t_\gamma \sim 10-100$ s), with
a non-thermal spectrum and varied on ms timescale, leads to the so-called
"compactness problem" (see Blinnikov 2000 for a deep physical discussion).
This ebergy liberated in a small region $\sim 10^6-10^7$ cm in size (as implied
by the ms variability time scale) would create a photon-lepton "fireball" with
enormous optical depth for pair creation by energetic photons, so a thermal
photon spectrum should be observed, unlike actually observed optically thin
non-thermal spectra. In addition, high-energy photons (with
$E_\gamma>10$~GeV) detected from some GRBs could not escape such a medium.

These problems can be circumvented if the fireball expands relativistically,
with a Lorentz-factor
$\Gamma>100-200$ (Ruderman 1975). Indeed, the size 
of the rapidly expanding volume, as derived from the emission time variability,
is $\Gamma^2$ as large as of the stable one. This and other relativistic
effects decreases the optical depth by a factor of $\Gamma^{7...8}$ or even
more (depending on geometry and other parameters),
 which solves the compactness problem of the fireball, but of cause
leaves open the question how such a fireball could be formed. But this is a
question to the "central engine" of GRBs.

Setting this explosion in tenuous interstellar (or intergalactic) medium results
in the formation of (collisionless) relativistic shocks (Rees and
M\'esz\'aros 1992). In this model, thermal energy of the initial
photon-lepton fireball with small baryon contamination ($\Delta M_b\sim
10^{-5} M_\odot$ to ensure a relativistic expansion with the required high
Lorentz factor) is transformed into the kinetic energy of baryons, which
sibsequently is thermalized in the relativistic shocks. It is this energy
that is eventually converted into X-ray photons (in the comoving frame) via
synchrotron and/or inverse Compton processes in the shocks, which are
detected as gamma-ray photons in the observer's frame.

In the currently most elaborated 
internal shock model (Narayan et al. 1992, 
Rees and M\'esz\'aros 1994), the GRB itself 
is produced when consecutive internal shocks, which 
are assumed to be generated by a (still unknown) "central engine" 
during the time $\Delta t_\gamma$  with slightly different Lorentz 
factors, collide with each other at a characteristic distance 
$r_c \sim \delta t c \Gamma^2\sim 10^{12}$~cm 
($\delta t\sim 10$ ms is a typical GRB variability time scale).
An external shock is formed at the collision site  
with the ambient medium ($\sim 10^{14}-10^{16}$ cm away from the
explosion site). The observed X-ray and optical afterglows
are thought to be produced by the external shocks when it decelerates
in the surrounding gas (in this respect GRB afterglows are just 
relativisic analogs of the conventional supernova remnants, i.e. 
are essentially environmental effects of powerful explosions).    

Basically, the model has only 6 free parameters: the initial energy of the
explosion $\Delta E$, the initial expansion Lorentz factor
$\Gamma$ (or, equivalently, the fireball baryonic load $\Delta M_b$), the
ambient gas density $n$, the fraction of the shock thermal energy in the
electronic component $\xi_e$, the fraction of the shock thermal energy
in the enhanced magnetic field energy behind the front $\xi_b$, and the
spectral index $p$ of accelerated relativistic electrons in shocks,
$dN/d\epsilon\sim \epsilon^{-p}$ (here $\epsilon$ is the electron energy).

The afterglow studies provide some evidence for possible beaming of
gamma-ray emission in GRBs (e.g. Frail et al. 2001, and
references therein). The angular beaming inferred is of order $\theta\sim
0.1$ rad. It is still unclear if a standard energy is released in 
GRBs or they have a broad luminosity function (the latter seems more
probable and seems to be required by extensive statistical
studies, e.g. Stern et al. 2001).  For
GRBs with known redshifts, $\Delta E$ can be directly derived from observed
flux assuming one or another GRB beaming factor. Typical values are
on average around $10^{53}$ ergs (assuming GRB emision isotropy), but
can be smaller by two orders of magnitude 
if the beaming factor (model-dependent estimate)
is taken into account (Lipunov et al 2001, Frail at al. 2001). 

Multiwavelength studies of GRB afterglows allow to determine 
the fireball model parameters, and show a broad consistence 
with GRB being superexplosions in the 
galactic environment (M\'esza\'aros 2001,
M\'esz\'aros and Rees 2001). 

\section{Central engine}

Much less is known about central engine of GRBs.
Small time-scale variability and large energy release suggest
the presence of a steallar-mass compact object. Gamma-ray 
beaming suggests the presence of rotating magnetized plasma.
Any viable mechanism for GRBs should be able to produce a relativistically
expanding fireball with small baryon contamination operating
during time intervals $\Delta T_{GRB} \sim 100$ s, which is 
much longer than the dynamical time scale for compact objects 
and the observed time variability scale (1-10 ms).

There is a growing evidence that GRBs are associated with 
star forming regions in galaxies (Piro et al. 2000, Sokolov et al. 2001),
so their progenitors could be massive stars ("failed supernovae" -- 
Woosley 1993, MacFadyen and Woosley 1999, 
"hypernovae" -- Pazcy\'nski 1997, "supranovae" -- 
Vietri et al. 2000,  WR-stars 
-- Postnov and Cherepashchuk 2001). 

Currently, 
two types of progenitors are considered: collapse of massive 
stars (as suggested by Woosley 1993), or coalescence of
binary compact objects (neutron stars or black holes) (as
first suggested by Blinnikov et al. 1984). Although strong
association of optically identified GRBs with star forming regions
favors the collapsar model, binary compact star coalescences
can not be totally discarded (for example, a significant
fraction of binary compact stars is expected to coalesce in a short time
after a star formation burst (Lipunov et al. 1995) so they, too,  
will be apparently associated with star forming regions in galaxies).
The collapsar model also suggest possible connection with 
supernova explosions, and indeed, apart from 
possible direct association of GRB980425, one can find some evidence
of this in optical afterglow behaviour monitoring (bumps in the
afterglow light curves, e.g. Sokolov 2001
and refs. therein), yet other explanations to these features
exist (Esin and Blandford 2000).

Certainly, the relativistic fireball model 
of GRBs have some problems (see e.g. Piran 2001) and
only new astronomical observations can resolve
all the GRB puzzles, but for our discussion here we will
rely upon the relativistic fireball model as mostly confirmed
by the existing observations. 

\section{High-energy phenomena associated with GRB}

The large energy release and relativistic expansion velocities 
implies that various high-energy phenomena can be associated with GRBs.
The relevant quantity derived from observations is 
the average energy production rate of cosmic GRBs. A simple estimate
is straightforward: the observed GRB rate is about ${\cal R}_{GRB}\sim 
10^{-7}$ per year per galaxy (assuming GRB isotropy), 
i.e. $\sim 10^{-9}$ per year per Mpc$^3$.
With the average isotropic energy of one GRB $10^{53}$ ergs we arrive 
at $dE/dt/dV\sim 10^{44}$ ergs per year per cubic Mpc.
Notably, this estimate does not depend on the GRB beaming
factor: with beaming, decrease in energy exactly compensates
for increase in rate.   

\subsection{Particle acceleration and ultra high-energy cosmic rays}

Fermi acceleration in relativistic internal 
shocks is likely to accelerate protons
to high Lorentz factors (Vietri 1995, Waxman 1995), so 
it is temptative to compare the above estimate with rough
energetics of ultra-high energy cosmic rays (see Waxman 2001 for extensive
discusion). Waxman shows  that ultra-high energy cosmic rays 
indeed could be produced by cosmological GRBs in the observed amount
(with unavoidable GZK-cutoff above $10^{19.7}$ eV).

According to this model, there are clear predictions 
potentially checkable by observations if GRBs indeed
provide a significant contribution to the observed flux of
UHECR. Namely, because of different time delay for protons 
with different energies in the intergalactic magnetic field,  
bursting UHECR sources should have narrowly
peaked spectra, with the brightest sources being different at different
energies. This feature is distinctive from steady sources
where the brightest sources at high energy should be brightest at
low energies as well. 
  
But in view of difficulties for {\it any} model involving 
population of sources at cosmological distances to explain 
the observed UHECR properties (especially the observed clustering
at different energies) (Teshima's lecture, this volume; 
see also a detailed discussion in Dubovsky 2001, this volume),
GRBs can hardly be considered as primaries for most energetic 
cosmic rays.

\subsection{Neutrino emission from GRBs}

In the framework of the relativistic fireball scenario,
GRB also become copious sources of high-energy neutrinos (Waxman and Bahcall
1997,1999).
Indeed, protons accelerated in the fireball lose energy via photo-meson 
interaction with fireball photons, mediated by $\Delta$-resonance
\footnote{If not only $\Delta$-approximation contributes to photo-meson
interaction, the results does not change significantly (Waxman 2001)}:
$p=\gamma\to \Delta \to p+\pi^+$. High energy neutrinos then 
result from the charged pions decays: $\pi^+\to \mu^++\nu_\mu\to
e^++\nu_e+\nu_\mu+\tilde\nu_\mu$. The relation between the
proton and neutrino energies is 
dictated by the $\Delta$-resonance condition $\epsilon_\gamma\epsilon_p=0.2
(\hbox{GeV}^2)\Gamma^2$. For typical $\epsilon_\gamma\sim 1$ MeV and 
$\Gamma\sim 300$ protons with energies $\epsilon_p\sim
10^{16}$ eV are capable of producing neutrinos. It can be shown that 
pion-decay produced neutrinos carry about 5\% of the proton energy, 
so the production of $\sim 10^{14}$ eV neutrinos are expected.

The GRB neutrino flux can be evaluated using observed gamma-ray fluences
$F_\gamma$.
Flux of pions is proportional to the proton flux, which is 
$f\pi\times F_\gamma/\xi_e$
Each neutrino carries $\sim 1/4$ of the pion energy, so the net result
is $\epsilon_\nu^2dN_\nu/d\epsilon_\nu\sim 0.25
(f_\pi/\xi_e)F_\gamma/\ln(10)$ (the factor $\ln 10$ accounts for the fact
that synchrotron radiating electrons span a decade in energies, 
as inferred from the observed GRB spectra). The average neutrino flux per
unit time
per solid angle is obtained by multiplying the single burst fluence
with the GRB rate per solid angle, which is $\sim 10^3$ bursts per year
per 4$\pi$ steradian. 
With a typical GRB 
fluence of $\sim 10^{-5}$ erg cm$^{-2}$ we arrive at a muon neutrino
flux $\epsilon_\nu^2\Phi_\nu\sim 3\times 10^{-9} (f_\pi/\xi_e)$
GeV~cm$^{-2}$s$^{-1}$sr$^{-1}$. Other neutrino flavors flux is
comparable.  
The expected high-energy muon neutrino flux can be detected by a km$^{2}$
neutrino detector with a rate of $\sim 10$ events correlated with GRBs.

The model allows the possibility 
to produce $\sim 10^{18}$ eV neutrinos in interaction of the reverse shock
 driven into the fireball ejecta at the initial stage of interaction of 
the fireball with the ambient medium. This takes place $\sim 10$ s after
the initial explosion, which is comparable with the fireball duration
itself. In this scenario, optical and UV-photons radiated by electrons
accelerated in shocks propagating backward into the ejecta may interact
with accelerated protons of the ejecta. A burst of $10^{17}-10^{19}$ eV 
neutrinos is then expected via photo-meson interactions (Waxman and Bahcall
2000). The estimated flux (somewhat more model-dependent) is
about $10^{-10}$GeV~cm$^{-2}$s$^{-1}$sr$^{-1}$, which can be more difficult
to detect.

Detection of high-energy neutrinos from GRB will test the shock 
acceleration mechanism and the key suggestion that GRBs are the sources
of the ultra-high energy protons ($>10^{16}$ eV to produce $>10^{14}$ eV 
neutrinos and $>10^{19}$ eV to produce $>10^{16}$ eV neutrinos).
   
In addition, inelastic $p-n$ collisions during the fireball acceleration
stage may produce $\sim 10$ GeV neutrinos with a fluence of $\sim 10^{-4}$
cm$^{-2}$ per burst. Such neutrinos can be potenially detected in a 1 km$^3$
neutrino telescope with a rate of $\sim 10$ events per year. There detection
will constrain the fireball neutron fraction and hence the GRB progenitor
model. 

\subsection{Gravitational waves production in GRBs}

The production of gravitational wave in GRB is much more model-dependent
since it require some model for the GRB progenitors. If compact binary
coalescences underly GRBs, gravitational waves will be copiously 
generated before
GRB, at the binary inspiral phase and at the merging phase. Binary
mergings are expected to be primary targets for the geravitaitonal
wave interferometers like LIGO, VIRGO or GEO-600, with the expected
event rate a few per year fro some types of compact binaries 
(see Grishchuk et al. 2001 for a recent review). Simultaneous
detection of GW signals from binary mergings with GRBs would
be a proof of the binary merging model for GRB. 

If GRBs are related to massive star core collapses, the situation is
less optimistic, since rather weak GW signals are expected to be associated
with collapses as both Newtonian and relativistic 
numerical simulations show (see 
Dimmelmeier et al. 2001 and references therein).

If relativistic fireball is beamed, as we discussed above,
there is a robust {\it lower} limit on the amount of gravitational
waves emitted due to the acceleration to relativistic
velocities of the fireball's baryons $m_b=\Delta E_\gamma/\Gamma c^2$
(Piran 2000). 
A broad-band signal is expected up to a maximal 
frequency $\omega_{max}\sim 2\pi/\delta t$, with a total amount of 
GW energy of $\Delta E_{GW}\sim 2Gm_b^2\Gamma^2/(c\pi\delta
t)$. At the maximum frequency $\nu_{max}\sim 100$ Hz this 
would correspond to
$$
h\sim Gm_b\Gamma^2/c^2d \approx 
3\times 10^{-25} (\Delta E_\gamma/10^{51}\hbox{erg})(\Gamma/100)
(d/100\,\hbox{Mpc})^{-1}\,,
$$
which is still too low for direct
detection even by the enhanced LIGO II interferometer.

The situation may be not so hopeless, however, if one 
appropriately uses correlation of GRB and GW (and possibly neutrino)
signals. At present, some joint data analysis algorithms
are under construction (e.g. Rudenko et al. 2000).

\subsection{GRB and theories of large and infinite extra dimensions}

A hugh energy reease in a compact region in GRB implies
very high energy densities resembling to certain stages 
in the radiation-dominated era of the evolution of the Universe.
With characteristic energy $E_{53}=\Delta E_\gamma/10^{53}$ erg
and the initial size $r_6=r_0/10^6$ cm the initial temperature
of the optically thick fireball is  
$$
T_f\simeq 116 (\hbox{MeV})E_{53}^{1/4}r_6^{-3/4}
$$
which is similar to that in the Universe as early as 
$\sim 10^{-4}$ s after the beginning of expansion. The photon number density
(as well as of relativistic leptons) is 
$$
n_\gamma\simeq 4.3\times 10^{37} (\hbox{cm}^{-3})(T/100\,\hbox{MeV})^3
$$ 
and diverse photon-photon and 
photon-lepton processes intensively occur. Thus the GRB fireballs
can be potentially useful to test high-energy physics at MeV scales.

We wish to consider constraints the very existence of GRBs
imposes on some modern theories of gravity. As example, we 
examine the theory of multi-dimensional gravity 
with quantum gravity scale at TeV energies
(Arkani-Hamed, Dimopoulos, Dvali 1998, hereafter ADD), and 
more recent 5D gravity of infinite-volume flat extra space
with $10^{-3}$ eV quantum gravity scale (Dvali, Gabadadze, Porrati
2000, Dvali et al. 2001, hereafter DGP). 
These theories assume that the Stabdard Model particles
are localized in a 3D "brane" embedded in compactified 
space with large (or infinitely large) extra dimensions.
The state-of-the-art in modern brane-world theories has been
extensively discussed in this conference (Yu.Kubyshin's talk), and in the
literature (Rubakov 2001, and references therein). 

In these theories, the fundamental gravity scale is no more
the conventional Planck mass  ($M_{P}\sim 10^{18}$ GeV),
which determines the observable 
weakness of the Newton gravitational constant $G_N$. 
The latter turns out to be 
defined by the quantum gravity scale $M_*$
of the corresponding theory.
In such a frame, one of the 
phenomenological manifestations of the existence
of large (or infinite) extra dimension(s) is an additional cooling
of hot plasma due to emission of Kaluza-Klein massive gravitons into 
the bulk (ADD model) or excitation of stringy Regge states (DGP model).

In the ADD scenario, the 4D Planck mass  
is related to the compactification radius $r_n$ and fundamental 
gravity scale $M_*$ as $M_P\sim r_n^n M_*^{n+2}$, where $n$ is the
number of extra dimensions. The emission of KK-gravitons in the bulk 
in photon-photon interactions (relevant to GRB fireballs) has a
cross section (Arkani-Hamed et al. 1998) 
$$
\sigma_{\gamma\gamma}\sim \frac{1}{16\pi}\left(\frac{T}{M_*}\right)^n
\frac{1}{M_*^2}
$$  
i.e. the KK-luminosity becomes 
$$
(dE/dt)_{KK}\sim n_\gamma^2\sigma_{\gamma\gamma}c\epsilon_{KK}
\propto T_f^{7+n}/M_*^{2+n}
$$  
(here $\epsilon_{KK}\sim 2.7 T_f$ is the typical KK-graviton energy).

If the emission of KK-gravitons effectively would cool down the
fireball before its initial thermal energy is converted into
the kinetic energy of the baryons, the required high
Lorentz factors would not be attained, and no GRB with 
the observed properties would be produced. This implies that
the emission of KK-gravitons in the fireball should meet
the condition
$$
r_0/c < \Delta E_\gamma/(dE/dt)_{KK}\,.
$$

Putting all quantities together, we arrive at the following constraints:
$$
n=2:\quad M_*>2(\hbox{TeV})E_{53}^{5/16}r_6^{-11/16}\,,
$$
$$
n=3:\quad M_*>0.25(\hbox{TeV})E_{53}^{3/10}r_6^{-7/10}\,.
$$
These are weaker than limits inferred from SN1987a neutrino
burst ($M_*>30$ TeV for n=2) and from cosmological considerations
(Arkani-Hamed et al. 1998; Hannestad, Raffelt 2001).

More interesting is the case of DGP model. In this framework, the weakness
of an observable gravity is explained by the high cut-off of the 
Standard Model $M_{SM}$ localized on the brane. 
In contrast to the ADD model, the large value of
the observable $M_P$ is determined my $M_{SM}\gg M_*$ 
rather than $M_*$. Now the emission of massive KK-gravitons into
the bulk is strongly suppressed. Instead, the possibility to
produce an exponentially large number of Regge states at very low
energy appears. At $T\ll 1$ TeV, the total rate of the production of stringy
Regge states is determined by the 2-d mass level and is 
(Dvali et al. 2001)
$$
\Gamma_2\sim E\frac{E^4}{M_*^2M_P^2} 
$$
(here the mean energy particle $E\sim 2.7 T$). 
This gives rise to the total Regge state emission rate in 
the GRB fireball
$$
(dE/dt)_R\approx 10^{55} (\hbox{erg/s})  E_{53}^{9/4}r_6^{-15/4}
$$
and the fireball acceleration constraints 
would be
$$
M_*>0.5\, (\hbox{eV}) E_{53}^{5/8}r_6^{-11/8}
$$
This is by about two orders of magnitude higher than 
original lower bound $10^{-3}$ eV discussed in Dvali et al (2001).
If this limit is true, deviations from the Newton gravity
are expected at distances smaller than 
$r<1/M_*\simeq 10^{-3}$ mm.   

\section{Conclusion}

After about 30 years of studies, the cosmic GRB phenomenon 
seems to be finally understood, at least in principle. 
It is an enormous energy release 
in a compact region, most plausibly due to either core collapse of a massive 
rotating star into a black hole, or binary compact star merging,
or both, in remote galaxies. 
This energy is in the form of photons and leptons,
with a small amount of baryons involved (a photon-lepton fireball).
The fireball rapidly expands, forming relativistic shock waves in the 
surrounding medium. The kinetic energy thermalized in the shocks 
produces the observed electromagnetic radiation of the GRB itself
and its afterglows in softer bands.
The nature of the central 
engine and generation mechanism of the (beamed) relativistic ejecta forming the
fireball remains  one of unresolved issues of the model.

In this framework, cosmic GRBs can be good accelerators of 
high-energy particles and thus contribute to the high-energy
cosmic rays. They also can be copious sources of high-energy
neutrinos, which can be detected in the forthcoming experiments.
If binary compact star mergings underly some GRBs, a time correlation with
chirp gravitational wave signals is expected, otherwise GRBs appear to 
be rather weak sources of gravitational waves. 

At last, a huge energy density in relativistic fireballs can be used as 
an independent tool to test modern theories of quantum gravity.

To conclude, we stress the need for both astrophysical multivawelength
studies of the GRB phenomenon (which proved to be extremely successful in 
the last years) and high-energy particle observations from GRBs.
These observations would provide independent tests for modern GRB theories
and can be used to study high-enegry particle properties. 

{\it Acknowledgements}. The author thanks V. Rubakov, Yu. Kubyshin, 
V. Rudenko, S. Dubovsky for useful discussions. Special thanks to 
D. Kosenko for help in some estimations. The work was partially
supported by RFBR grants 99-02-16205, 00-02-17884a, and 00-02-17164.

\end{document}